\documentstyle[psfig]{article}

\begin{document}

\title{Efficiency of Brownian Motors}
\author{J.M.R. Parrondo\\
{\small \em Dept. de F\'{\i}sica
At\'omica, Nuclear y Molecular, Universidad Complutense,
28040-Madrid, Spain}\\[.2cm]
J.M. Blanco, F.J. Cao, and R. Brito\\
{\small \em Dept. de F\'{\i}sica
Aplicada I, Universidad Complutense,
28040-Madrid, Spain}}

\maketitle

Pacs: 05.40,82.20M

\begin{abstract}
The efficiency of different types of Brownian
motors is calculated analytically and numerically.
We find that motors based on flashing
ratchets present a low efficiency and an unavoidable
entropy production.
On the other hand, a certain class of
motors based on adiabatically changing potentials, named
{\em reversible ratchets},
exhibit a higher efficiency and the entropy production can
be arbitrarily reduced.
\end{abstract}

\begin{figure}
\[
\psfig{figure=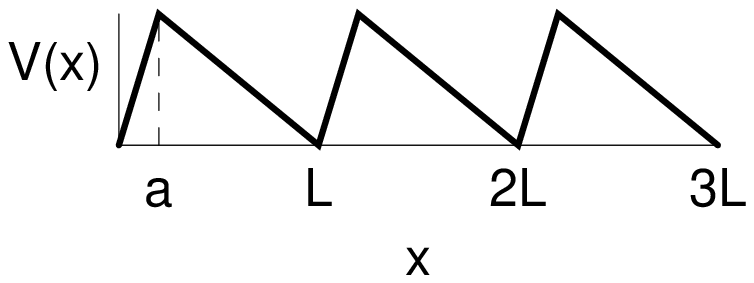,height=2.5cm}
\hspace{1cm}
\psfig{figure=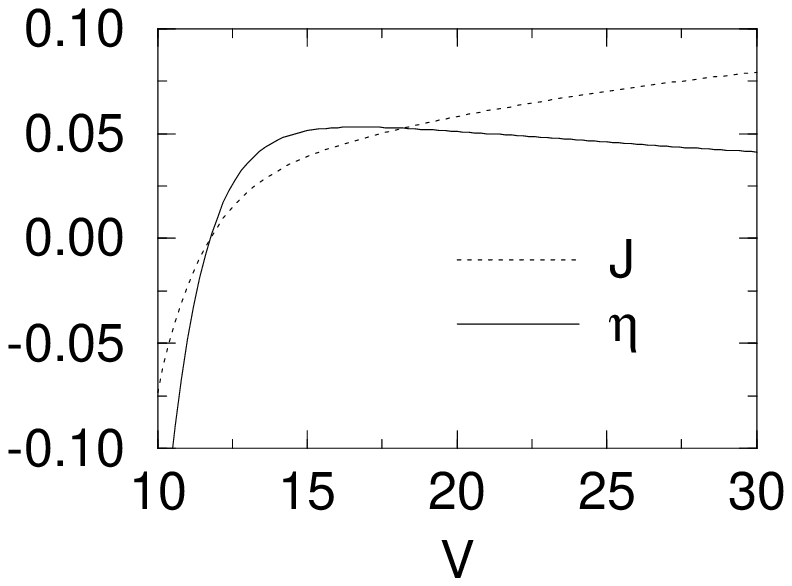,width=5cm}
\]
\caption[]{{\em Left}:
Asymmetric sawtooth potential of the ratchets
presented in refs. 
\cite{mag1,prost,astumian,doe,hang}. 
In this letter we consider two types of
ratchets: a)  one where the
potential is randomly switched on and off; and b) one
where the potential is deterministically modulated.
\\
{\em Right}: Efficiency and current of the ratchet where the
the potential on the left 
is randomly switched on and off (case a),
as a function of the maximum height $V$ of the potential.
The reaction
rates are $\omega_A=1.08$ and $\omega_B=81.8$, $a=1/11$,
 and the
external force is $F=4.145$.}
\label{fig-mag}
\end{figure}

In the last years there has been an increasing
interest in the so-called
``ratchets'' or Brownian motors 
\cite{mag1,prost,astumian,doe,hang,prost2}. 
These systems consist
of Brownian particles moving in
asymmetric potentials, such as the one depicted in
fig.\ \ref{fig-mag} (left), 
and subject to a source of
non-equilibrium,
like external fluctuations or temperature gradients.
As a consequence of
these two ingredients ---asymmetric potentials and
non-equilibrium---, a 
flow of particles can be induced.

Most of the cited papers consider systems where
the Brownian particles do not gain energy in
a systematic way. Although these systems are 
called ``Brownian or molecular motors'', they do not convert
heat into work, nor induce any energy conversion.
Feynman in his {\em Lectures} \cite{feyn}
already understood that, in order
to have an engine out of a ratchet, it is necessary to use
its systematic motion to  store potential energy.
This can be achieved if the ratchet lifts a load.
Then the ratchet becomes a thermal engine and Feynman
estimated its efficiency (although following 
assumptions which
have been revealed to contain some inconsistencies
\cite{parr}).
Recently, Sekimoto \cite{sekimoto} has generalized this
procedure, defining efficiency for a wide class
of ratchets. J\"ulicher {\it et al} \cite{prost2}
have also discussed the efficiency of 
molecular motors and
Sokolov and Blumen \cite{sok} have calculated the
efficiency of a deterministically flashing
ratchet in contact
with thermal baths at different temperatures.
A general conclusion is that  these motors are 
intrinsically irreversible, even in the quasistatic limit
\cite{prost2,parr,sekimoto,sok}.

On the other hand, it has been recently introduced
\cite{parr2}
a class of deterministically
driven ratchets where the entropy production vanishes
in the quasistatic limit, {\it i.e.},
{\em reversible ratchets}.
The aim of this letter is to explore the
differences, regarding efficiency, between
randomly flashing ratchets and both
reversible and irreversible deterministically driven
ratchets.  

\subsubsection*{Randomly flashing ratchets}
Consider two species of Brownian 
particles, say $A$ and $B$, moving
in the interval $[0,L]$ with periodic boundary
conditions. 
Particles of type $A$ feel a potential $V_A(x)$, whereas particles
$B$ 
feel $V_B(x)$. Besides, there is a continuous transfer of 
particles,
$A\rightleftharpoons B$, which accounts for non-equilibrium 
fluctuations. This picture is equivalent to that
of a single Brownian particle in
a randomly switching potential \cite{astumian}.

In ref. \cite{astumian}, it was proved that a flow towards a given
direction, say, to the right,
 occurs for some asymmetric potentials $V_A$ and $V_B$.
If we add a load or force $F$ opposite to the
flow, the evolution equation for the probability
density of particles $A$, $\rho_A(x)$, and particles $B$,
$\rho_B(x)$, is:
\begin{eqnarray}\label{fp1}
\partial_t \rho_A(x,t) & =& -\partial_x {\cal J}_A\rho_A(x,t)-
\omega [ \rho_A(x,t)-\rho_B(x,t)]\nonumber\\
\partial_t \rho_B(x,t) & =& -\partial_x {\cal J}_B\rho_B(x,t)+
\omega [ \rho_A(x,t)-\rho_B(x,t)]
\end{eqnarray}
where ${\cal J}_{i}=-V_i'(x)-F-\partial_x$ is the
{\em current operator},
the
prime indicates derivative with respect to $x$, and $\omega$
is the rate of the reaction 
$A\rightleftharpoons B$.
We have taken
units of energy, length and time 
such that the temperature is $k_BT=1$, the
length of the interval is $L=1$, and
the diffusion coefficient is
$D=1$. 

The flow of particles in the stationary regime
is $ J={\cal J}_{A}\rho^{st}_A(x)+
{\cal J}_{B}\rho^{st}_B(x)$,
where  $\rho_{A,B}^{st}(x)$ are the stationary
solutions of eq. (\ref{fp1}).
This flow $J$ is a decreasing function of the external
force $F$
and  becomes negative if $F$ is stronger than a
{\em balancing force}, $F_{bal}$. 
Therefore, if $0<F<F_{bal}$, particles move against the force
and, consequently, gain potential energy in a systematic
way. The potential energy
gain or {\em output energy} per unit of time is
$E_{out} = JF$, which vanishes both for
$F=0$ and $F=F_{bal}$. 

On the other side, switching on and off the potential requires
some energy. In our two-species picture, the reaction 
$A\rightleftharpoons B$ does not conserve energy since
$V_A(x)\ne V_B(x)$. Therefore, in each reaction $A\to B$,
occurring at a point $x$, an energy 
$V_B(x)-V_A(x)$ is transferred to the system (or withdrawn, if
the sign is negative). Similarly,  an energy
$V_A(x)-V_B(x)$ is transferred to the system in each
reaction $B\to A$ occurring at $x$. In the
stationary regime, the average number of such reactions
per unit of time is, respectively, $\omega\rho^{st}_A(x)$
and $\omega\rho^{st}_B(x)$. Therefore, the 
{\em input energy} per unit of time is \cite{prost2,sekimoto}:
\begin{equation}
E_{in}= \omega\int_0^1 dx\ \left[V_B(x)-V_A(x)\right]
\left[\rho^{st}_A(x)-\rho^{st}_B(x)\right].
\end{equation}
Finally, the efficiency can be defined 
as \cite{prost2,sekimoto}:
\begin{equation}
\label{eff1}
\eta={E_{out}\over E_{in}}.
\end{equation}

This efficiency can be calculated analytically for
the system given by eq. (\ref{fp1}) with piecewise
potentials. We have performed an exhaustive study for
the particular setting $V_B(x)=0$ and  $V_A(x)$ 
equal
to the potential depicted in fig. \ref{fig-mag} (left):
\begin{equation}
\label{potm}
V_A(x) =  \left\{
\begin{array}{ll}
Vx/a & \mbox{if $x\le a$} \\
V(1-x)/(1-a) & \mbox{if $x\ge a$} 
\end{array}
\right.
\end{equation}
with $a=1/11$.
For this system, the maximum efficiency 
is $\eta_{max}=3.29\%$,
which is reached for
 $V=22$, $F=3$, and $\omega=63$. This efficiency can
be improved using different reaction rates:
$\omega_A$ for $A\to B$ and $\omega_B$ for $B\to A$.
In this case, $\eta_{max}=5.315\%$ with $V=16.7$, $F=4.145$,
$\omega_A=1.08$, and $\omega_B=81.8$. Observe that,
with these values for $\omega_{A,B}$, the particle
stays much longer within the potential $V_A(x)$ than 
within $V_B(x)$. 

We have
plotted in fig. \ref{fig-mag} (right)
the efficiency and the flow of particles
as a function of $V$ setting the rest of 
parameters equal to these optimal values.
Two are the messages from this figure.
Firstly, the maximization of the efficiency is a
new criterion to define optimal Brownian motors and this
criterion is, in some cases, less trivial than that of
maximizing the flow.
Secondly,
the randomly flashing ratchet under study 
has a rather low efficiency.
As we have mentioned before, the heat dissipation
per unit of time is $E_{in}-E_{out}$.
Consequently, the increase of entropy
of the thermal bath, per unit of time, is
$E_{in}-E_{out}$, since $k_BT=1$.
On the other hand, in the stationary regime
there is no change of entropy in
the system nor in the external agent which provides
the non-equilibrium fluctuations\footnote{
A physical realization of this external agent is
a third species of particles, say $C$, feeling a potential
$V_C(x)=V_B(x)-V_A(x)$ and participating in the reaction as
$A+C\rightleftharpoons B$.
If the temperature of $C$ particles is the same
as $B$ and $A$ particles, then detailed balance holds and
there is no flow of particles. 
However, if the temperature of the $C$
particles is infinity, we recover the flashing ratchet
discussed in the text. Therefore, this randomly
flashing ratchet can be
considered as a thermal engine in contact with two thermal baths,
one at $T=1/k_B$ and the other one at infinite temperature
(see also \cite{sok} for an interpretation of
the deterministically flashing ratchet as a thermal engine).}.
Therefore, the net entropy production per unit of time
is $E_{in}-E_{out}$.
If this entropy production
would vanish, {\it i.e.}, if the system could work in a
reversible way,
it should reach a 100\% efficiency. On the contrary, 
the efficiency is below 10\% and we can 
conclude that the motor based on the randomly 
flashing ratchet is very inefficient.


One could think that the efficiency
would increase in limiting
situations where the system is close to equilibrium, 
such us $\omega\to 0$ and/or $V_A-V_B\to 0$.
However, a perturbative analysis of eq.  (\ref{fp1}) shows
that $\eta\to 0$ in both limits.
In the first case, $\omega\to 0$,
from eq. (\ref{fp1}) one can easily find that $J$
is of order $\omega$, so is $F_{bal}$. Therefore,
$E_{out}$, in the interval $0<F<F_{bal}$, is of
order $\omega^2$, whereas one can prove that $E_{in}$
is of order $\omega$, giving a zero efficiency in this limit.
In the second case,
$\Delta V(x)\equiv V_A(x)
-V_B(x)\to 0$, the input energy $E_{in}$ is of order
$\Delta V^2$. However, surprisingly enough, $J$ is
of order $\Delta V^2$ and so is $F_{bal}$, yielding
$E_{out}$ of order $\Delta V^4$ and, again, a 
vanishing efficiency. 
We conclude that the flashing motor is intrinsically irreversible,
as it has been pointed out for related
models in refs. \cite{prost2,parr,sekimoto,sok}.

\subsubsection*{Deterministically driven ratchets}
A different strategy to reduce the production of entropy
consists of considering Brownian particles in a 
potential which changes
deterministically in time. If the potential is changed very slowly,
the system evolves close to equilibrium and the 
entropy production is low.
From now on, we focus
our attention on
Brownian particles in a spatially periodic 
potential $V(x;{\bf R}(t))$ 
depending on a set of parameters collected in a
vector ${\bf R}$ which changes in time \cite{parr2}. 
The parameters are changed periodically in time with
period $T$, {\it i.e.}, ${\bf R}(0)= {\bf R}(T)$.

As in ref. \cite{sekimoto},
we have to modify
our definition of efficiency. Firstly, we deal with
energy transfer per cycle $[0,T]$ instead per unit of
time. Secondly, the input energy 
or work done to the system in a cycle,
as a consequence of the change
of the parameters ${\bf R}(t)$, is:
\begin{equation}
\label{ein1}
E_{in}= \int_0^T dt {\partial V(x;{\bf R}(t))\over
\partial t}\rho(x,t).
\end{equation}
The probability density $\rho(x,t)$ verifies the
Smoluchowski equation:
\begin{equation}
\label{eq-smol}
\partial_t\rho(x,t) =-\partial_x{\cal J}_{{\bf R}(t)}
\rho(x,t)
\end{equation}
where ${\cal J}_{\bf R}= -V'(x;{\bf R})-F-\partial_x$ is
the current operator corresponding to the potential
$V(x;{\bf R})$. As before, the output energy is the current
times the force $F$,
but now the current is not stationary and we have
to integrate along the process:
\begin{equation}
\label{eout1}
E_{out}=\int_0^T dt\ F{\cal J}_{{\bf R}(t)}\rho(x,t)=F\phi.
\end{equation}
where $\phi$ is the integrated flow.

With the above expressions,
the efficiency of the system, $\eta=E_{out}/E_{in}$,
 can be found analytically
for $T$ large  and weak external forces, where
it is expected to be high.  For the integrated
flow one finds 
$\phi=\phi_0-\bar\mu FT$, where
 $\bar\mu$ is the average mobility of
the system:
\begin{equation}
\label{mob}
\bar\mu={1\over T}\int_0^T {dt\over Z_+({\bf R}(t)) 
 Z_-({\bf R}(t))}
\end{equation}
and $\phi_0$ is the integrated flow for $F=0$ \cite{parr2}:
\begin{equation}
\label{phi0}
\phi_0=\oint d{\bf R}\cdot
 \int_0^1dx\int_0^xdx' \rho_+(x;{\bf R})
\nabla_{\bf R}\rho_-(x';{\bf R}),
\end{equation}
with  
\[
\rho_{\pm}(x;{\bf R})={e^{\pm V(x;{\bf R})}\over Z_\pm({\bf R})};
\quad
Z_{\pm}({\bf R})=\int_0^1dx\ e^{\pm V(x;{\bf R})}.
\]
In eq. (\ref{phi0}) 
 the contour integral runs over the closed
path $\{{\bf R}(t):t\in [0,T]\}$ in the space of parameters
of the potential.
The term proportional to $T$ in the integrated
flow, $\phi=\phi_0-\bar\mu FT$, arises because
the force $F$ induces a non-zero current which is present
along the whole process. As a consequence,
the balancing force is 
$F_{bal}=\phi_0/\bar\mu T$, and, in order to design a
high efficiency motor, it is necessary to
take simultaneously  the adiabatic limit
$T\to \infty$ and the limit $F\to 0$ with $ FT$
finite. Notice also that the above expressions
are useless if $\phi_0=0$. 
In a previous paper \cite{parr2}, we have
discussed the conditions for $\phi_0$ to be different from zero
and called {\em reversible ratchets} those systems 
where $\phi_0\ne 0$.
From now on, we restrict our analytical calculations to reversible
ratchets, although we present below  numerical results for
an irreversible ratchet.

The input energy for weak force $F$ and large $T$ is 
$ E_{in}=\phi_0F+b/ T$, with
\begin{eqnarray}
b &=& -\int_0^T dt\ Z_-({\bf R}(t))Z_+({\bf R}(t))\left\{
\left[ \int_0^1dx \int_0^xdx'\ \rho_+(x;{\bf R}(t))
\left[\partial_t\rho_-(x';{\bf R}(t))\right]
\right]^2\right. \nonumber \\
&+& 
\left.\int_0^1dx \int_0^xdx'
 \int_0^{x'}dx''
\ \left[\partial_t\rho_-(x;{\bf R}(t)) \right]
\ \rho_+(x';{\bf R}(t))
\left[\partial_t\rho_-(x'';{\bf R}(t)) \right] 
\right\}
\end{eqnarray}
which is a positive quantity.
Combining the above expressions, one finds for the efficiency:
\begin{equation}
\label{efficiency}
\eta = 
{F(\phi_0-\bar\mu FT)\over \phi_0F+b/T}=
{\phi_0 \alpha -\bar\mu \alpha^2\over \phi_0\alpha +b}
\end{equation}
where $\alpha=FT$. This expression is exact
in the limit $T\to \infty$, $F\to 0$.
Notice that, even for large $T$, 
the irreversible contribution, $b/T$, to
$E_{in}$ is of the same order as $\phi_0F$.

In a given system, {\it i.e.}, for a set
of parameters $\phi_0$, $\bar\mu$ and $b$, the maximum efficiency
is reached for 
$\alpha=(b/\phi_0)[\sqrt{1+\phi_0^2/(\bar\mu b)}-1]$
and its value is given by
\begin{equation}
\label{maxeff}
\eta_{max}= 1-2\left[\sqrt{z(1+z)}-z\right]
\end{equation}
with $z=b\bar\mu/\phi^2_0$.
Eq. (\ref{maxeff}) clearly shows how the 
term $b$ in the denominator 
of eq. (\ref{efficiency}) prevents the system to reach an 
efficiency equal to one.  Fortunately, as we will
see below in a concrete example, using strong
potentials one can 
get arbitrarily
close to 100\% efficiency.

To  check the validity of the above theory and
stress the differences between reversible and irreversible
ratchets, we have studied in detail one example of
each class.

As an example of irreversible ratchet, consider the
modulation of the 
potential in fig. \ref{fig-mag} (left),
{\it i.e.}, $V(x;t)= \cos^2(\pi t/T)V(x)$
with $V(x)$ given by eq. (\ref{potm}). 
In this case, $\phi_0$ is zero and the above theory
cannot be applied. 
We have numerically
integrated the Smoluchowski equation, eq. (\ref{eq-smol}), 
using an
implicit scheme with $\Delta t=10^{-5}$, $\Delta x=0.002$, $0.005$,
and the Richardson extrapolation method to correct inaccuracies
coming from the finite $\Delta x$. The efficiency has been 
obtained using eqs. (\ref{eff1}), (\ref{ein1}), 
(\ref{eout1}) and the results, as a function of $F$
and for different values of $T$, are plotted in fig. 
\ref{fig-eff} (left).
The efficiency is maximum for $T$ around
0.5 and it goes to zero as $T$ increases. 
The maximum efficiency
found by numerical integration
is of the same order of magnitude as the one found for
the randomly flashing ratchet. Notice, however, that 
we cannot explore  with numerical experiments the whole space
of parameters.

\begin{figure}[t]
\[
\psfig{figure=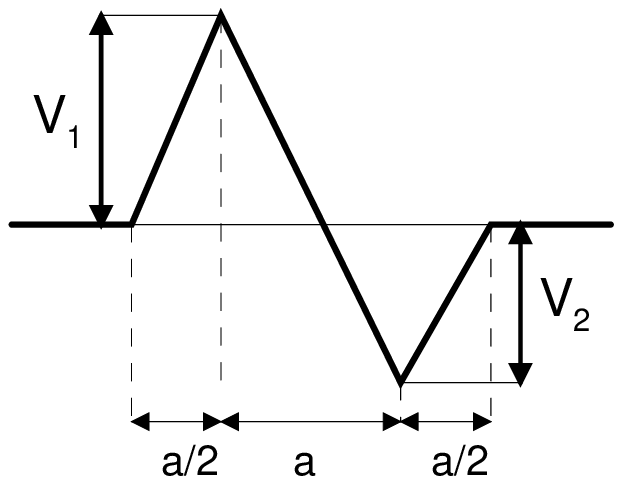,width=3.6cm}
\hspace{1cm}
\psfig{figure=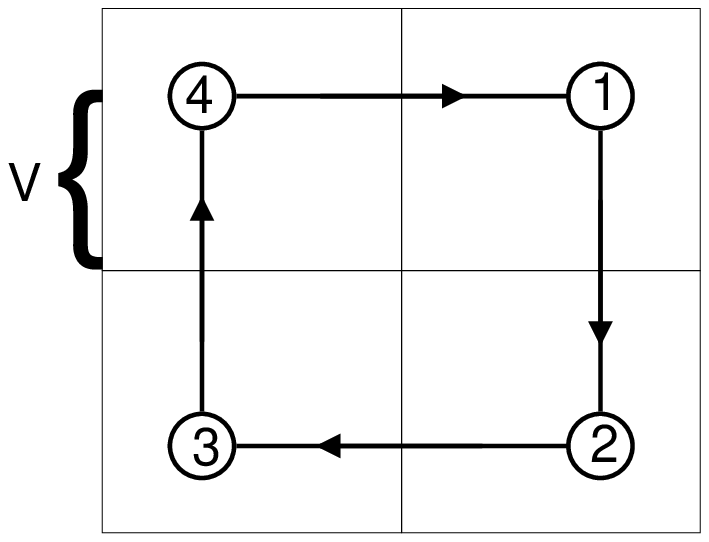,width=3.6cm}
\hspace{1cm}
\psfig{figure=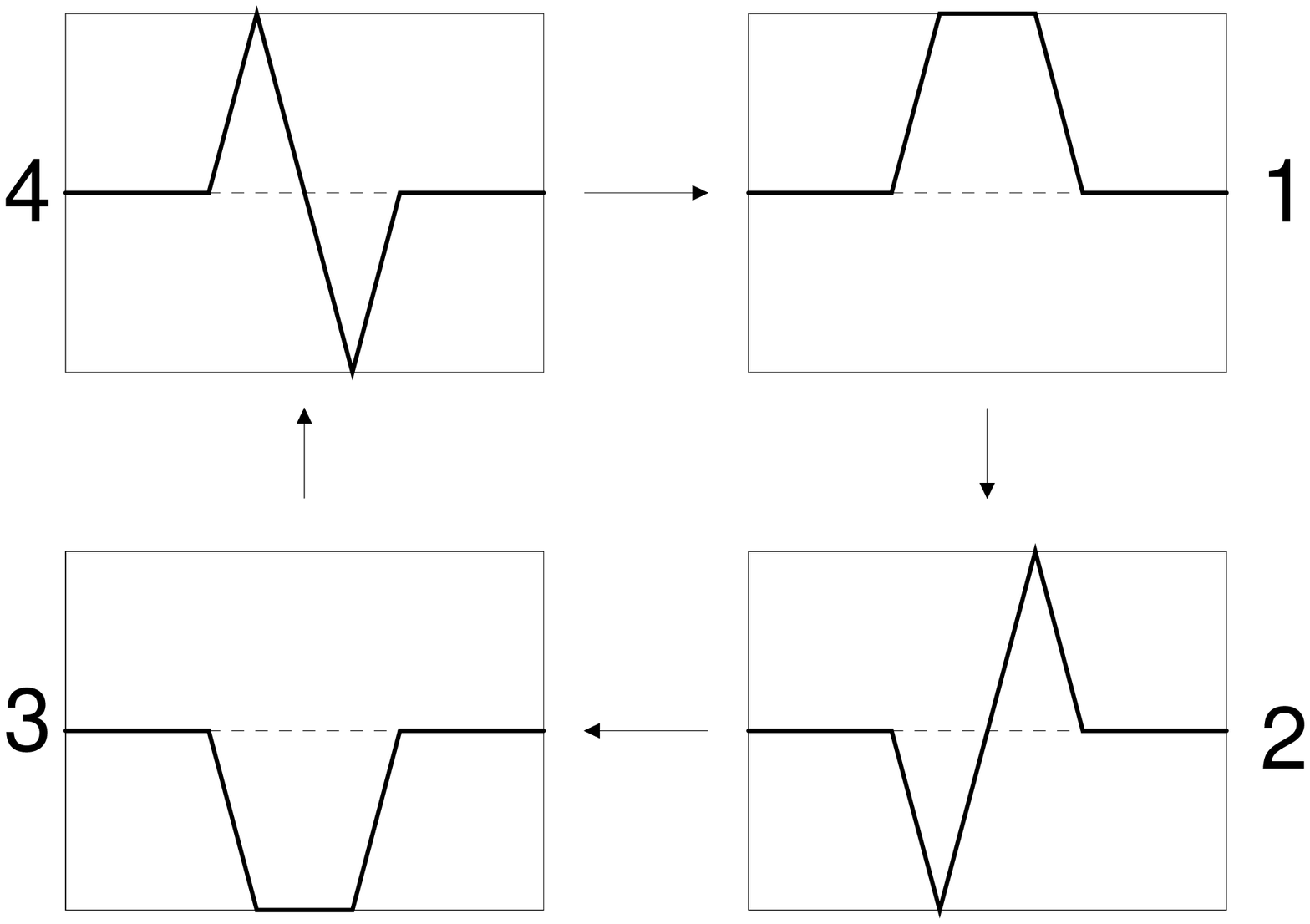,width=4.0cm}
\]
\caption[]{Graphical representation of the
reversible ratchet described in the text: the potential
depends on two parameters, $V_1$ and $V_2$, which
are the height of two barriers/wells (left) and
they change along the path depicted on
the center ($V$ being the maximum height/depth
of the barriers/wells). On the right, the shape
of the potential at the four labelled points.
}
\label{fig-potrev}
\end{figure}

\begin{figure}[h]
\[
\psfig{figure=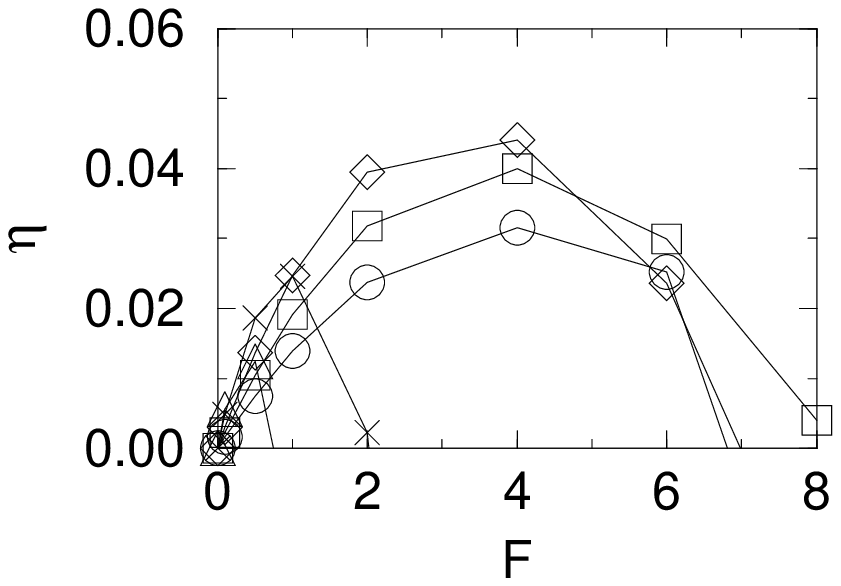,width=4.3cm}
\hspace{1cm}
\psfig{figure=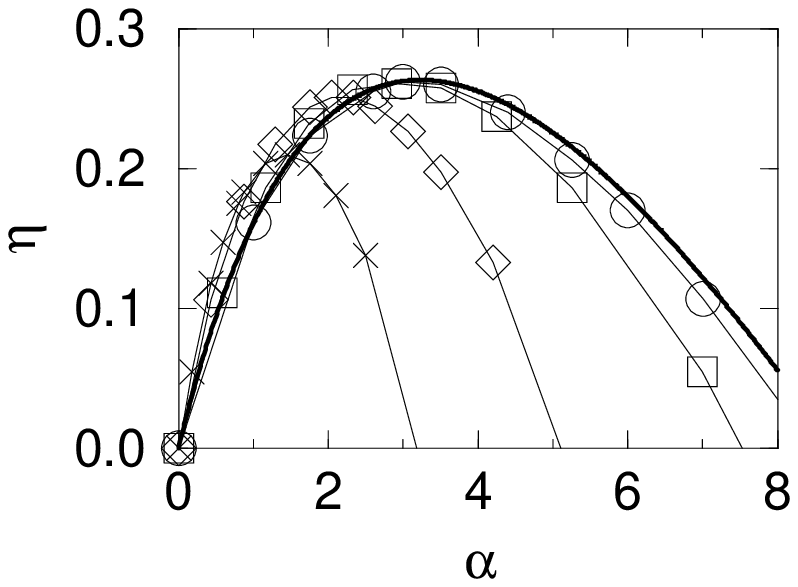,width=4cm}
\]
\caption[]{
{\em Left (irreversible ratchet)}: numerical
results for the 
efficiency of the ratchet consisting of the potential
in fig. \ref{fig-mag}  (left)
modulated by $z(t)=\cos^2(\pi t/T)$
as a function of the external force $F$ and for different
values of the period $T$: $T$= 0.00125 ($\bigcirc$), 0.025
($\Box$), 0.05 ($\Diamond$), 0.25 ($\times$), and
0.5 ($\triangle$).
\\
{\em Right (reversible ratchet)}: numerical and analytical 
results for the
efficiency of the ratchet described
in fig. \ref{fig-potrev} for $V=5$ $a=0.2$ as a
function of $F$ and for different values of the period $T$:
$T$= 1 ($\times$), 2
($\Diamond$), 10 ($\Box$), 40 ($\bigcirc$).
The thick solid line is the analytical 
result given by eq. (\ref{efficiency})
in the limit $T\to\infty$ and $F\to 0$.
Note that $\eta$ is an increasing function of $T$ in
the reversible ratchet (right) as opposite to the irreversible
case (left).
}
\label{fig-eff}
\end{figure}

On the other hand,  let us consider the reversible ratchet 
represented in fig. \ref{fig-potrev}. Here the potential
depends on two parameters, $V_1$ and $V_2$, which are
the heights/depths of two triangular barriers/wells of
width $a$. 
The ratchet consists of modifying at constant
velocity the parameters $V_1$ and $V_2$ 
along the path depicted in the same figure (center).
This example is
a modification of the one presented in ref. \cite{parr2}.
Now $\phi_0$ does not vanish and the above theory gives us 
the efficiency in the limit $T\to\infty$ and $F\to 0$. For
instance, for $V=5$ and $a=0.2$, we obtain $\phi_0=0.825$,
$\bar\mu=0.094$ and $b=3.74$. The efficiency given 
by eq. (\ref{efficiency}) is plotted in fig.
 \ref{fig-eff} (right) and it is compared 
 with the numerical integration
of the Smoluchowski equation for different values of $T$. Notice
the differences with the irreversible ratchet. Here the 
efficiency is an increasing function of $T$. 
The maximum efficiency, for the parameters corresponding
to fig. \ref{fig-eff}, is 26\% and
is almost reached for $T=40$.
The efficiency of this ratchet can be arbitrarily
close to 100\% if one increases $V$.
The reason is that the average mobility
decreases exponentially with $V$, but the coefficient $b$ and
the integrated flow $\phi_0$ remain finite. 
For instance, for $V=20$ and $a=0.4$, $\phi_0=0.999988$,
$b=6.89$ and $\bar\mu<10^{-7}$, yielding
a maximum efficiency of 99.85\%.

To summarize, we have calculated
the efficiency of a randomly flashing ratchet 
with an asymmetric sawtooth potential. 
In order to find more efficient Brownian motors, 
we have also calculated the efficiency
of deterministically driven ratchets, finding that
the efficiency of reversible ratchets is
much higher than the efficiency of irreversible ratchets.
It is remarkable that
the class of reversible ratchets 
involves potentials depending on two or 
more parameters \cite{parr2}
and they differ from
the models considered to date in the literature. 
Here we have shown that this new and non trivial
class of ratchets is a real 
breakthrough  regarding efficiency.

This work has been financially supported by
Direcci\'on General de Investigaci\'on Cientifica y
T\'ecnica (DGICYT)
(Spain) Project No. PB94-0265.
J.M.B. and F.J.C. acknowledge financial support 
from the program {\em Becas de Colaboraci\'on} of
the Ministerio de Educaci\'on (Spain).

\end{document}